\documentstyle[12pt,epsf]{article}
\begin{document}
\title{{\bf Fluctuation Kinetics in a Multispecies
Reaction-Diffusion System}}
\author{Martin Howard}
\date{
{\small\noindent{\it Department of Physics, Theoretical Physics,
University of Oxford, 1 Keble Road, Oxford, OX1 3NP, United Kingdom.}
\\}
}\maketitle
\begin{abstract}
We study fluctuation effects in a two species reaction-diffusion
system, with three competing reactions $A+A\rightarrow\emptyset$,
$B+B\rightarrow\emptyset$, and $A+B\rightarrow\emptyset$.
Asymptotic density decay rates are calculated for $d\leq 2$ using two
separate methods - the Smoluchowski approximation, and also field
theoretic/renormalisation group (RG) techniques.
Both approaches predict power law decays, with exponents which
asymptotically depend
only on the ratio of diffusion constants, and not on the reaction
rates. Furthermore, we find that, for $d<2$, the Smoluchowski
approximation and the RG improved tree level give identical exponents.
However, whereas the Smoluchowski approach cannot easily be improved,
we show that the RG provides a systematic method for incorporating
additional fluctuation effects. We demonstrate this advantage by
evaluating one loop corrections for the exponents in $d<2$, and find
good agreement with simulations and exact results.

\
\begin{center}
\end{center}
\end{abstract}
\newpage
\section{Introduction}

Over the past decade there has been enormous interest in
reaction-diffusion
systems (see [1--12] and references therein), with particular emphasis
on the effects of
fluctuations in low spatial dimensions. Most attention has been paid
to reactions of the form $A+A\rightarrow\emptyset$ and
$A+B\rightarrow\emptyset$ with a variety of different initial/boundary
conditions. At or below an upper critical dimension $d_c$, these
systems exhibit fluctuation induced anomalous kinetics, and the
straightforward application of
traditional approaches, such as mean field rate equations, breaks
down. Attempts to understand the role played by fluctuations for
$d\leq d_c$ have involved several techniques, including Smoluchowski
type approximations \cite{KBR} and field theoretic methods
\cite{L,LC,HC}. In this paper we set out to study these fluctuation
effects in a system with three competing reactions:
$$
A+A\rightarrow\emptyset \qquad B+B\rightarrow\emptyset \qquad
A+B\rightarrow\emptyset.
$$
The reactions are irreversible, and we choose homogeneous, though not
necessarily equal, initial densities for the two species at $t=0$. Our
goal is to calculate density decay exponents and amplitudes,
taking into account fluctuation effects. In pursuit of this
aim, we analyse the system using both the Smoluchowski approximation
and the field theory approach, and we show that the two methods are
closely related. However, whereas it is unclear how the Smoluchowski
approach may be improved, the field theory provides a systematic way
to obtain successively more accurate values for the asymptotic density
decay
exponents and amplitudes. We shall concentrate on situations where one
of the two species is greatly in the majority (as is almost always the
case asymptotically) - so, for example, if
the A species is predominant, then we can safely neglect the reaction
$B+B\rightarrow\emptyset$. This kind of assumption will lead to a
considerable simplification in our analysis.

Previous work on this problem includes use of the Smoluchowski
approximation \cite{KBR}, as well as exact $1d$ results obtained by
Derrida {\it et al.} \cite{D1,D2,D3} for the special case
of {\it immobile} minority particles. Derrida {\it et al.}
were, in fact, studying a different problem, namely the probability
that a given spin has never flipped in the zero temperature Glauber
dynamics
of the q-state Potts model in one dimension. By solving that model
exactly \cite{D2,D3}
they showed that this probability decreased as a power law: $t^{-3/8}$
for the Ising ($q=2$) case. However, in one dimension, the Ising spin
flip problem and
the decay rate for the immobile impurity in our reaction-diffusion
system are exactly equivalent problems, and hence this exact
decay rate
also holds in our case. We also mention one other previous result for
the immobile impurity problem, due to Cardy \cite{C}. Using
renormalisation
group methods similar to those employed in this paper, it was shown
that the density of the minority species decays away as a universal
power law: $t^{-\beta}$ for $d<2$, where $\beta={1\over
2}+O(\epsilon)$ and $\epsilon=2-d$.

The case where the {\it majority} species is immobile has also been
solved (see \cite{PG}). In this case the decay
rate for the minority
species is dominated by minority impurity particles existing in
regions where
there happen to be very few of the majority particles. Since these
majority particles are strictly stationary, this
situation is not describable using a rate equation approach, and it
turns out that the minority species decays away as
$\exp\left(-t^{d/(d+2)}\right)$, a result which is not accessible by
perturbative methods.

In this paper, using a field theory formalism and techniques
 from the renormalisation group, we will obtain decay rates and
amplitudes for the general case of arbitrary diffusivities - a regime
previously only accessible using the Smoluchowski approximation. Our
basic plan is to map the microscopic dynamics, as described by a
master equation, onto a quantum field theory. This theory is then
renormalised (for $d\leq 2$), and the couplings (reaction rates) are
shown to have $O(\epsilon)$ fixed points, whose values depend only on
the ratio of the species' diffusion constants. Note that this system
(with irreversible reactions) is particularly simple in that
only the couplings (and not the diffusivities) are renormalised. The
next step is to group together Feynman diagrams which are of the same
order in the renormalised couplings - i.e. diagrams with the same
number of loops. These diagrams are then evaluated and a
Callan-Symanzik equation used to obtain improved asymptotic
$\epsilon$ expansions for the densities. In this fashion, quantities
of interest may be systematically calculated by successively including
higher order sets of diagrams (with more loops) in the perturbative
sum.

One consequence of the theory is that the asymptotic decay
rates and
amplitudes for $d<d_c$ will be independent of the reaction rates
- a result which is in accordance with the Smoluchowski approach. In
fact, all physical quantities below the upper critical dimension
asymptotically depend only on the diffusivities and the initial
densities, and in this sense they display universality.

We now present a summary of our results for the density decay
rates. In what follows we define $n_A$, $n_B$ to be the initial
density of A, B particles, and $\delta=(D_B/D_A)\leq 1$ to be the
ratio of the diffusion constants.
For $d<d_c=2$, $n_A\gg n_B$ and
$n_A^{-2/d}D_A^{-1}\ll t\ll t_{1}$ (where $t_{1}$ is a crossover
time derived in section 4), we have (as in \cite{L}):
\begin{equation}
\langle a\rangle\sim\left({1\over 4\pi\epsilon}+{2\ln 8\pi-5\over
16\pi}+O(\epsilon)\right) (D_At)^{-d/2}.
\end{equation}
For the minority species, we find, from the RG improved tree level
approximation in the field theory:
\begin{equation}
\langle b\rangle\sim F(D_At)^{-\beta}
\end{equation}
where
\begin{equation}
\beta\approx{d\over 2}\left({\delta +1\over 2}\right)^{d/2} \qquad
F\approx n_B\left({\Gamma(\epsilon/2)\over n_A(8\pi)^{d/2}}
\right)^{\left({\delta+1\over 2}\right)^{d/2}}.
\end{equation}
These decay exponents are identical with the Smoluchowski
results. Performing a strict $\epsilon$ expansion on this RG improved
tree level result gives an exponent $\beta={1\over 2}+O(\epsilon)$
for the immobile impurity case ($\delta=0$). This is in agreement with
previous RG calculations by Cardy \cite{C}. If we now go beyond the
tree level calculation by including one loop diagrams, then we obtain
an improved value for the exponent $\beta$ using an $\epsilon$
expansion:
$$
\beta=\left({1+\delta\over 2}\right)\left(1-{\epsilon\over
2}\left[{3\over 2}+\ln\left({1+\delta\over 2}\right)-
{\delta(1+\delta)\over 4}\left[1+2\ln\left({1+\delta\over
2}\right)\right]
\right.\right.
$$
\begin{equation}
\left.\left. -{1\over
4}(\delta^2-1)\left(1+(1+\delta)\left[f\left\{{2\over
1+\delta}\right\}-{\pi^2\over 6}\right]\right)\right]\right)
+O(\epsilon^2),
\end{equation}
where
\begin{equation}
f\{x\}=-\int_1^x{\ln u\over u-1}du
\end{equation}
is the dilogarithmic function \cite{AS}. This exponent is found to be
in good agreement with simulations \cite{KBR} and exact results
\cite{D2} in $d=1$.

However, for $\delta<1$,
the system crosses over to a second regime where $\langle
b\rangle\gg \langle a\rangle$. This situation is similar to the case
where we begin with $n_B\gg n_A$. In that regime, at times $D_Bt\gg
n_B^{-2/d}$, and for $n_B\gg n_A$, $\delta\neq 0$ and $d<2$, we have:
\begin{equation}
\langle b\rangle\sim\left({1\over 4\pi\epsilon}+{2\ln 8\pi-5\over
16\pi}+O(\epsilon)\right)(D_Bt)^{-d/2},
\end{equation}
for the majority species. Using the RG improved tree level result for
the minority species, we obtain:
\begin{equation}
\langle a\rangle\sim E(D_Bt)^{-\alpha},
\end{equation}
with
\begin{equation}
\alpha\approx{d\over 2}\left({1+\delta^{-1}\over 2}\right)^{d/2}
\qquad E\approx n_A \left({\Gamma(\epsilon/2)\over
n_B(8\pi)^{d/2}}\right)^{\left({1+\delta^{-1}\over 2}\right)^{d/2}}.
\end{equation}
\def\gaeq{>\mkern 4mu\sim}
The exponent is again in agreement with the Smoluchowski result. If we
attempt to improve this calculation to one loop accuracy, then we
obtain:
$$
\alpha=\left({1+\delta^{-1}\over 2}\right)\left(1-{\epsilon\over
2}\left[{3\over 2}+\ln\left({1+\delta^{-1}\over 2}\right)-
{\delta^{-1}(1+\delta^{-1})\over
4}\left[1+2\ln\left({1+\delta^{-1}\over 2}\right)\right]
\right.\right.
$$
\begin{equation}
\left.\left. -{1\over
4}(\delta^{-2}-1)\left(1+(1+\delta^{-1})\left[f\left\{{2\over
1+\delta^{-1}}\right\}-{\pi^2\over 6}\right]\right)\right]\right)
+O(\epsilon^2).
\end{equation}
This exponent is only valid for $\delta$ quite close to unity, and
even in this region it may be less accurate
than the (non $\epsilon$-expanded) RG improved tree level result given
above. This point will be discussed further in section 4.2.

We next give results valid for $d=2$, where we find extra
logarithmic factors multiplying the power law decay rates. Treating
first the case $\langle a\rangle\gg \langle b\rangle$ and $\delta\leq
1$, we have, from the RG
improved tree level, an initial regime with:
\begin{equation}
\langle a\rangle\sim{\ln t\over 8\pi D_At}
\end{equation}
\begin{equation}
\langle b\rangle =O\left(\left({\ln t\over
t}\right)^{\left({1+\delta\over 2}\right)}\right).
\end{equation}
However, for $\delta<1$, the system again crosses over to a
second regime where $\langle b\rangle\gg\langle a\rangle$. In this
second regime the density decay exponents (though not the amplitudes)
are the same as for the case where we begin with $n_B\gg n_A$.
In that case we have, for $\delta\neq 0$:
\begin{equation}
\langle b\rangle\sim{\ln t\over 8\pi D_Bt}
\end{equation}
\begin{equation}
\langle a\rangle =O\left(\left({\ln t\over
t}\right)^{\left({1+\delta^{-1}\over 2}\right)}\right).
\end{equation}
Crossover times for these cases are given in section 4.3.

We now give a brief description of the layout of this paper. In the
next section we analyse the system using the mean field/Smoluchowski
approach. We then set up the necessary formalism for our field theory
in section 3, and use it to perturbatively calculate values
for the density exponents and amplitudes in section 4. Finally, we
give some conclusions and prospects for future work in section 5.

\section{The Mean Field and Smoluchowski Approach}

The simplest description of a
reaction-diffusion process is provided by the mean field rate
equations. For the system we are considering with densities $a$ and
$b$, they take the form:
\begin{eqnarray}
& & {da\over dt}=-2\lambda_{AA}a^2-\lambda_{AB}ab \label{ratea} \\
& & {db\over dt}=-2\lambda_{BB}b^2-\lambda_{AB}ab, \label{rateb}
\end{eqnarray}
where $\lambda_{AA}$, $\lambda_{BB}$, and $\lambda_{AB}$ are the
reaction rates, and where we impose initial conditions of the form
$a|_{t=0}=n_A$ and $b|_{t=0}=n_B$.
In this approach we have completely neglected the effects of
fluctuations - in other words we have made assumptions of the form
$\langle ab\rangle\propto\langle a\rangle\langle b\rangle$ etc., where
the angular brackets denote averages over the noise. Below the
critical dimension, where fluctuations become relevant, this
sort of approximation will break down.

Nevertheless, even at the mean field level, the complete solution set
for these rate equations is quite complicated. In what follows we
shall restrict our analysis to the case where
$2\lambda_{BB}<\lambda_{AB}<2\lambda_{AA}$. The solution for this
particular parameter set will be required for our later field
theoretic analysis. Following \cite{KBR}, it is easy to show (by
forming a rate equation for the concentration ratio)
that $(a/b)\rightarrow 0$ as $t\rightarrow\infty$. Thus if we begin
with initial conditions where $n_A\gg n_B$, we can identify two
distinct regimes - an early time regime where $a\gg b$ and, after a
crossover, a late time (true asymptotic) regime where $b\gg a$.
Treating the early time regime first, we find (after some algebra):
\begin{eqnarray}
& & a\sim(2\lambda_{AA}t)^{-1} \label{mf1a}\\
& & b\sim{n_B\over(2n_A\lambda_{AA}t)^{\lambda_{AB}/2\lambda_{AA}}}.
\label{mf1b}
\end{eqnarray}
Note that the A
particles are decaying away more quickly than the B's, so eventually
we crossover to a second regime:
\begin{eqnarray}
& & b\sim(2\lambda_{BB}t)^{-1} \label{mf2a} \\
& & a\sim{n_A\over (2n_B\lambda_{BB}t)^{\lambda_{AB}/2\lambda_{BB}}}
\left(1+{(\lambda_{AB}-2\lambda_{AA})\over
(2\lambda_{BB}-\lambda_{AB})}
{n_A\over n_B}\right)^{-1-{\lambda_{AB}(\lambda_{AB}-2\lambda_{BB})
\over 2\lambda_{BB}(\lambda_{AB}-2\lambda_{AA})}}. \label{mf2b}
\end{eqnarray}
Alternatively, if we begin with $n_B\gg n_A$,
then we have a single asymptotic regime:
\begin{eqnarray}
& & b\sim(2\lambda_{BB}t)^{-1} \label{mf3a}\\
& & a\sim{n_A\over
(2n_B\lambda_{BB}t)^{\lambda_{AB}/2\lambda_{BB}}}. \label{mf3b}
\end{eqnarray}

However, if we now wish to extend our results at or below the upper
critical dimension, we must attempt
to include some of the fluctuation effects. The simplest way in which
this can be done is to employ the Smoluchowski approximation
\cite{SM,CH,KBR}. The essential idea of this approach is to relate the
effective reaction rates $\lambda^{eff}_{\{ij\}}$ to the diffusion
constants
$D_A,D_B$. Suppose we want to calculate the reaction rate
$\lambda^{eff}_{AB}$. We begin by choosing a (fixed) A species target
``trap'', which is
surrounded by B particles. When a B particle approaches within a
distance $R$ of the target, a reaction is deemed to have
occurred. Consequently, the reaction rate may be obtained by solving a
diffusion equation with boundary conditions of fixed density as
$r\rightarrow\infty$, and absorption at $r=R$. The flux of B
particles across the $d$ dimensional sphere of radius $R$ is then
proportional to an
effective microscopic reaction rate. If we now generalise to the case
where both the A and B species are mobile, then we find (in dimension
$d<2$ and in the large time limit):
\begin{equation}
\lambda^{eff}_{AB}\sim ({\rm const.})(D_A+D_B)^{d/2}t^{d/2-1}.
\end{equation}
For $d=2$ we obtain logarithmic corrections:
\begin{equation}
\lambda^{eff}_{AB}\sim {({\rm const.})(D_A+D_B)\over\ln((D_A+D_B)t)}.
\end{equation}
The Smoluchowski reaction rates for $\lambda^{eff}_{AA}$ and
$\lambda^{eff}_{BB}$
are obtained in a similar fashion. Note that
above $d=2$ the reaction rate approaches a limiting (constant) value,
and we see that the Smoluchowski approach predicts a critical
dimension of $d_c=2$ for this system. This is simply related to the
reentrancy property of random walks in $d\leq 2$. It is the
inclusion of this effect which accounts for the
improvement introduced by the Smoluchowski approach.

If we now substitute these modified reaction rates into the rate
equations, we can obtain the Smoluchowski
improved density exponents. For the case where $n_A\gg
n_B$, we find an initial regime with
\begin{eqnarray}
& & a=O(t^{-d/2}) \\
& & b=O\left(t^{-{d\over 2}\left({1+\delta\over
2}\right)^{d/2}}\right).
\end{eqnarray}
Once again, since the A particles are decaying away
faster than the B's, we cross over to a second regime, where (for
$0<\delta<1$)
\begin{eqnarray}
& & b=O(t^{-d/2}) \\
& & a=O\left(t^{-{d\over 2}\left({1+\delta^{-1}\over
2}\right)^{d/2}}\right).
\end{eqnarray}
This second set of exponents is the same as for the case where
we begin with $n_B\gg n_A$ and $\delta\neq 0$. In this situation no
crossover occurs and the exponents are valid for all asymptotic times.
These exponents can be compared
favourably with both simulations \cite{KBR}, and exact results
\cite{D3}. For example, the decay rate for an immobile minority
impurity is given by Smoluchowski to be $\approx t^{-0.354}$. This
compares well with the exact decay rate of $t^{-0.375}$.

Turning to the case $d=d_c=2$ and $n_A\gg n_B$, we obtain, for the
initial regime:
\begin{eqnarray}
& & a=O\left({\ln t\over t}\right) \\
& & b=O\left(\left({\ln t\over t}\right)^{\left({1+\delta\over
2}\right)}(\ln t)^{\left({1+\delta\over
2}\right)\ln\left({1+\delta\over 2}\right)}\right).
\end{eqnarray}
We again eventually crossover to a second regime, where (for
$0<\delta<1$):
\begin{eqnarray}
& & b=O\left({\ln t\over t}\right) \\
& & a=O\left(\left({\ln t\over t}\right)^{\left({1+\delta^{-1}\over
2}\right)}(\ln t)^{\left({1+\delta^{-1}\over
2}\right)\ln\left({1+\delta^{-1}\over 2}\right)}\right).
\end{eqnarray}
This second set of exponents is again valid (for all asymptotic times)
in the case where we begin with $n_B\gg n_A$ and $\delta\neq 0$.

Note that the Smoluchowski approach can also be employed for $d>d_c$,
where again we will find (time independent) reaction rates which
depend on the diffusion constants. However, in the general case, our
later field theoretic analysis shows that there
is no real justification for this procedure. One exception to this
occurs in the case where we have heterogeneous single species
annihilation, as considered in \cite{KBR}. In this situation we
have only one fundamental reaction process, but different reaction
rates may still arise, for example, by having two or more
different particle masses (and hence two or more different diffusion
constants). In this case it is physically
reasonable to suppose that the exponents (which are ratios of reaction
rates) may depend only
on the diffusivity ratios, with any other parameters canceling
out. However, in the general case, where the
reaction processes are genuinely distinct this will not be the case.

Overall, we have seen that the Smoluchowski approach is a simple way
to incorporate some fluctuation effects into the rate equation
approach. Unfortunately, it is not at all clear how these methods may
be systematically improved. It is for this reason that we turn to the
main purpose of this paper - the development of an alternative field
theoretic framework.

\section{The Field Theory Approach}

Fluctuation effects in reaction-diffusion systems have previously
been successfully tackled using techniques borrowed from quantum field
theory and also from the renormalisation group. Examples include
studies of the diffusion limited reactions $A+A\rightarrow\emptyset$
\cite{L} and $A+B\rightarrow\emptyset$ \cite{LC,HC}. The first step in
this analysis is to write down a Master
Equation, which exactly describes the microscopic time evolution of
the system. Using methods developed by Doi \cite{Doi} and Peliti
\cite{Pel}, this can be mapped onto a Schroedinger-like
equation, with the introduction of a second quantised
Hamiltonian, and then onto a
field theory, with an action $S$. These steps have been described in
detail elsewhere \cite{Doi,Pel,L,LC,HC}, and consequently
we shall simply give the resulting action appropriate for our theory:
\begin{eqnarray}
& & S=\int d^dx \left( \int dt \left[\bar a
(\partial_t-\nabla^2)a+\bar b
(\partial_t-\delta\nabla^2)b
+2\lambda_{AA}\bar aa^2+\lambda_{AA}\bar a^2a^2
\right.\right. \label{action} \\ & &
\left.\left. +2\lambda_{BB}\bar
bb^2+\lambda_{BB}\bar b^2b^2+\lambda_{AB}\bar aab+\lambda_{AB}\bar
bab+\lambda_{AB}\bar a\bar bab \right] -\bar an_A -\bar
bn_B\right). \nonumber
\end{eqnarray}
Here we have defined $\delta=(D_B/D_A)\leq 1$ and also introduced the
response fields $\bar a$ and $\bar b$. In
addition time $t$, together with the
reaction rates $\lambda_{\{ij\}}$ have been rescaled to absorb the
diffusion constant $D_A$. Averaged quantities are then calculated
according to
\begin{equation}
\langle X(t)\rangle={\cal N}^{-1}\int {\cal D}a\,{\cal D}
\bar a\,{\cal D}b\,{\cal D}\bar b\,X(t)\, e^{-S},
\end{equation}
where
\begin{equation}
{\cal N}= \int{\cal D}a\,{\cal D}\bar a\,{\cal
D}b\,{\cal D}\bar b\, e^{-S}.
\end{equation}
Notice that in the path integral
\begin{equation}
\int{\cal D}a\,{\cal D}\bar a\,{\cal D}b\,{\cal D}\bar b\, e^{-S}
\end{equation}
integration over the fields $\bar a$, $a$ and $\bar b$, $b$
whilst neglecting the quartic terms, leads to a recovery of the mean
field rate equations.

Performing power counting on the action $S$, we can now give the
natural canonical dimensions for the various parameters appearing in
the action:
\begin{equation}
[t]\sim k^{-2}\qquad [a],[b],[n_A],[n_B]\sim k^d \qquad [\bar a],[\bar
b]\sim k^0 \qquad [\lambda_{\{ij\}}]\sim k^{2-d}.
\end{equation}
Notice that the reaction rates become dimensionless in $d=2$, which we
therefore postulate as the upper critical dimension for the system,
in agreement with the Smoluchowski prediction.

{}From the action $S$, we can see that the propagators for the theory
are given by
\begin{eqnarray}
& & G_{a\bar a}(k,t-t')=\cases{e^{-k^2(t-t')} & for $t>t'$\cr
0 & for $t<t'$\cr} \\
& & G_{b\bar b}(k,t-t')=\cases{e^{-k^2(t-t')\delta} & for $t>t'$\cr
0 & for $t<t'$.\cr}
\end{eqnarray}
Diagrammatically, we represent $G_{a\bar a}$ by a thin solid line and
$G_{b\bar b}$ by a thin dotted line. The vertices for the theory are
given in figure 1.

\subsection{Renormalisation}

One of the most important features of this theory, as mentioned in the
introduction, is the relative simplicity of its
renormalisation. Examination of the vertices given in figure 1
reveals that it is not possible to draw diagrams which dress
the propagators. Hence the bare propagators are the full propagators
for the theory. Consequently, the only renormalisation needed involves
the reaction rates $\lambda_{\{ij\}}$, and in particular the diffusion
constants (or $\delta$) are {\it not} renormalised.

The temporally extended vertex functions for the reaction rates are
given by the diagrammatic sums given in figure 2. As is the case in
similar theories \cite{L,LC,HC}, these sums may be evaluated exactly,
using Laplace transforms:
\begin{eqnarray}
& & \lambda_{AA}(k,s)={\lambda_{AA}\over
1+\lambda_{AA}C\Gamma(\epsilon/2)(s+{1\over 2}k^2)^{-\epsilon/2}}
\label{tevf1} \\
& & \lambda_{BB}(k,s)={\lambda_{BB}\over
1+\lambda_{BB}C\Gamma(\epsilon/2)\delta^{-1}(s/\delta +{1\over
2}k^2)^{-\epsilon/2}} \label{tevf2} \\
& & \lambda_{AB}(k,s)={\lambda_{AB}\over
1+\lambda_{AB}2^{-\epsilon/2}C\Gamma(\epsilon/2)(1+\delta)^{-d/2}(s+
k^2\delta/(1+\delta))^{-\epsilon/2}} \label{tevf3},
\end{eqnarray}
where $C=2/(8\pi)^{d/2}$ and $s$ is the Laplace transformed time
variable.

We can now use these vertex functions to define the three
dimensionless renormalised and bare
couplings, with $s=\kappa^2$, $k=0$ as the normalisation point:
\begin{equation}
g_{R_{\{ij\}}}=\kappa^{-\epsilon}\lambda_{\{ij\}}
(k,s)|_{s=\kappa^2,k=0}\qquad\quad
g_{0_{\{ij\}}}=\kappa^{-\epsilon}\lambda_{\{ij\}}.
\end{equation}
Consequently, we can define three $\beta$ functions:
\begin{eqnarray}
& & \beta(g_{R_{AA}})=\kappa{\partial\over\partial\kappa}g_{R_{AA}} =
-\epsilon g_{R_{AA}}+\epsilon C\Gamma(\epsilon/2)g_{R_{AA}}^2
\label{b1} \\
& & \beta(g_{R_{BB}})=\kappa{\partial\over\partial\kappa}g_{R_{BB}} =
-\epsilon g_{R_{BB}}+\epsilon C\Gamma(\epsilon/2)\delta^{-d/2}
g_{R_{BB}}^2 \label{b2} \\
& & \beta(g_{R_{AB}})=\kappa{\partial\over\partial\kappa}g_{R_{AB}} =
-\epsilon g_{R_{AB}}+2^{-\epsilon/2}
\epsilon C\Gamma(\epsilon/2)(1+\delta)^{-d/2} g_{R_{AB}}^2, \label{b3}
\end{eqnarray}
and three fixed points $\beta(g_{R_{\{ij\}}}^*)=0$:
\begin{eqnarray}
& & g_{R_{AA}}^*=(C\Gamma(\epsilon/2))^{-1} \\
& & g_{R_{BB}}^*=(C\Gamma(\epsilon/2)\delta^{-d/2})^{-1} \\
& & g_{R_{AB}}^*=\left(C\Gamma(\epsilon/2){1\over 2}
\left({1+\delta\over 2}\right)^{-d/2}\right)^{-1}.
\end{eqnarray}
Finally, we see from (\ref{tevf1}), (\ref{tevf2}), and (\ref{tevf3})
that the expansion of $g_{0_{\{ij\}}}$ in powers of $g_{R_{\{ij\}}}$
is given by:
\begin{equation}
g_{0_{\{ij\}}}=g_{R_{\{ij\}}}+{g_{R_{\{ij\}}}^2\over g_{R_{\{ij\}}}^*}
+\ldots \label{gexp}
\end{equation}

\subsection{Callan-Symanzik Equation}

We now exploit the fact that physical quantities calculated using the
field theory must be independent of the choice of normalisation
point. This leads us to a Callan-Symanzik equation:
\begin{equation}
\left[\kappa{\partial\over\partial\kappa}+\beta(g_{R_{AA}}){\partial
\over\partial g_{R_{AA}}}+\beta(g_{R_{BB}}){\partial
\over\partial g_{R_{BB}}}+\beta(g_{R_{AB}}){\partial
\over\partial g_{R_{AB}}}\right]\langle
a\rangle_R=0.
\end{equation}
However dimensional analysis implies
\begin{equation}
\left[\kappa{\partial\over\partial\kappa}-2t{\partial\over\partial
t}+dn_A{\partial\over\partial n_A}+dn_B{\partial\over\partial
n_B}-d\right]\langle
a\rangle_R(t,n_A,n_B,g_{R_{\{ij\}}},\delta,\kappa)=0.
\end{equation}
Exactly similar equations hold for $\langle b\rangle_R$. Eliminating
the terms involving $\kappa$ and solving by the method of
characteristics, we find:
\begin{equation}
\langle a\rangle_R(t,n_A,n_B,g_{R_{\{ij\}}},\delta,\kappa)=(\kappa^2
t)^{-d/2}\langle a\rangle_R(\kappa^{-2},\tilde n_A(\kappa^{-2}),\tilde
n_B(\kappa^{-2}),\tilde g_{R_{\{ij\}}}(\kappa^{-2}),\delta,\kappa),
\label{CSS}
\end{equation}
with the characteristic equations:
\begin{equation}
2t{\partial\tilde n_A\over\partial t}=-d\tilde n_A \qquad
2t{\partial\tilde n_B\over\partial t}=-d\tilde n_B \qquad
2t{\partial\tilde g_{R_{\{ij\}}}\over\partial t}=\beta(\tilde
g_{R_{\{ij\}}}), \label{gchar}
\end{equation}
and initial conditions:
\begin{equation}
\tilde n_A(t)=n_A \quad \tilde n_B(t)=n_B
\end{equation}
\begin{equation}
\tilde g_{R_{AA}}(t)=g_{R_{AA}} \quad
\tilde g_{R_{BB}}(t)=g_{R_{BB}} \quad
\tilde g_{R_{AB}}(t)=g_{R_{AB}}.
\end{equation}
These equations have the exact solutions:
\begin{equation}
\tilde n_A(t')=\left({t\over t'}\right)^{d/2}n_A \qquad \tilde
n_B(t')=\left({t\over t'}\right)^{d/2}n_B, \label{RD}
\end{equation}
and
\begin{equation}
\tilde g_{R_{\{ij\}}}(t')=g^*_{R_{\{ij\}}}\left(1+{g^*_{R_{\{ij\}}}-
g_{R_{\{ij\}}}\over
g_{R_{\{ij\}}}(t/t')^{\epsilon/2}}\right)^{-1}.
\label{RC}
\end{equation}
In the large $t$ limit $\tilde g_{R_{\{ij\}}}\rightarrow
g^*_{R_{\{ij\}}}$, a relationship which will allow us to relate an
expansion in powers of the renormalised couplings $g_{R_{\{ij\}}}$ to
an $\epsilon$ expansion using (\ref{CSS}). In our later density
calculations we will assume that this asymptotic regime has been
reached.

\subsection{Tree Diagrams}

In order to perform systematic $\epsilon$ expansion calculations
we now need to identify the leading and subleading terms in an
expansion in powers of $g_{0_{\{ij\}}}$. In calculating $\langle
a\rangle$ and
$\langle b\rangle$, contributions from tree diagrams are of order
$g_{0_{\{ij\}}}^q n_{\{i\}}^{1+q}$, for integer
$q$, and densities $n_{\{i\}}=\{n_A,n_B\}$. However, diagrams with $l$
loops
will be of order $g_{0_{\{ij\}}}^{q+l} n_{\{i\}}^{1+q}$. The addition
of loops makes the power $g_{0_{\{ij\}}}$ higher
relative to the power of the densities - so we conclude that the
number of loops gives the order of the diagram.

The lowest order diagrams contributing to $\langle a\rangle$ and
$\langle b\rangle$ are the tree diagrams shown in figure 3. We
represent the classical (tree level) density $\langle a\rangle_{cl}$
by a wavy solid line, and $\langle b\rangle_{cl}$ by a wavy dotted
line. These sets of diagrams are equivalent to the mean field
rate equations, as may be seen by acting on each by their respective
inverse Green functions.

The second tree level quantities appearing in the theory are the
response functions:
\begin{eqnarray}
& & L(k,t_2,t_1)=\langle a(-k,t_2)\bar a(k,t_1)\rangle \\
& & M(k,t_2,t_1)=\langle b(-k,t_2)\bar a(k,t_1)\rangle \\
& & N(k,t_2,t_1)=\langle b(-k,t_2)\bar b(k,t_1)\rangle \\
& & P(k,t_2,t_1)=\langle a(-k,t_2)\bar b(k,t_1)\rangle
\end{eqnarray}
which we represent diagrammatically by the thick lines shown in figure
4. These functions can be evaluated analytically, but only in the
limit $\langle a\rangle\gg \langle b\rangle$, or $\langle b\rangle\gg
\langle a\rangle$. The details of this calculation are
presented in appendix A, where the following results are derived (for
$\langle a\rangle\gg \langle b\rangle$):
\begin{eqnarray}
& & L(k,t_2,t_1)=\left({1+2\lambda_{AA}n_At_1\over
1+2\lambda_{AA}n_At_2}\right)^2\exp{(-k^2(t_2-t_1))}\label{R1L}\\
& & N(k,t_2,t_1)=\left({1+2\lambda_{AA}n_At_1\over
1+2\lambda_{AA}n_At_2}\right)^{\lambda_{AB}/2\lambda_{AA}}
\exp{(-k^2(t_2-t_1)\delta)}\label{R1N}\\
& & P(k,t_2,t_1)=-\lambda_{AB}n_A{(1+2\lambda_{AA}n_At_1)^{
\lambda_{AB}/2\lambda_{AA}}\over(1+2\lambda_{AA}n_At_2)^2}\exp{(-k^2
(t_2-t_1\delta))}\label{R1P} \nonumber \\ & & \qquad
\qquad\qquad\qquad\times\int_{t_1}^{t_2}{\exp{(k^2(1-\delta)t')}\over
(1+2\lambda_{AA}n_At')^{-1+\lambda_{AB}/2\lambda_{AA}}}dt' \\
& & M(k,t_2,t_1)=-\lambda_{AB}n_B{(1+2\lambda_{AA}n_At_1)^2
\over(1+2\lambda_{AA}n_At_2)^{\lambda_{AB}/2\lambda_{AA}}}\exp{(-k^2
(t_2\delta-t_1))} \nonumber \\ & & \qquad
\qquad\qquad\qquad\times\int_{t_1}^{t_2}{\exp{(-k^2(1-\delta)t')}\over
(1+2\lambda_{AA}n_At')^{2}}dt'. \label{R1M}
\end{eqnarray}
An extra check on validity of these response functions is provided by
the relations:
\begin{eqnarray}
& & L(0,t,0)={\partial\langle a(t)\rangle\over\partial n_A} \qquad
N(0,t,0)={\partial\langle b(t)\rangle\over\partial n_B} \\
& & P(0,t,0)={\partial\langle a(t)\rangle\over\partial n_B} \qquad
M(0,t,0)={\partial\langle b(t)\rangle\over\partial n_A},
\end{eqnarray}
which follow from the definition of the response functions, and from
the initial condition terms in the action $S$. It is easy to check
that the above response functions do indeed satisfy these relations.

For the opposite situation where $n_B\gg n_A$ (and hence $\langle
b\rangle\gg\langle a\rangle$), we could
use a formalism similar to the above for the density
calculations. However, it is
much simpler to map this case onto the $\langle
a\rangle\gg\langle b\rangle$ regime by
swapping the labels on the A and B particles, and
then relabeling:
$$
n_A\leftrightarrow n_B \quad
\lambda_{AA}\leftrightarrow\lambda_{BB} \quad D_A\leftrightarrow D_B.
$$
We can then obtain the exponents and amplitudes for this
second regime with no extra work.

This concludes our discussion of the field theory formalism. The
framework we have built up allows (in principle) the systematic
calculation of fluctuation effects in all circumstances. However, it
is only in the case where one of the species is greatly in the
majority where the equations (for the tree level densities and
response functions) are sufficiently simple for analytic progress to
be made. We now turn to use of the field theory in calculating the
fluctuation modified densities.

\section{Density Calculations}

\subsection{Tree Level}

The first step in using our field theory to include fluctuation
effects is to insert the mean field (tree level) solution into the
Callan-Symanzik solution (\ref{CSS}), using the results for the
running densities/couplings (\ref{RD}), (\ref{RC}). Since the fixed
points for the couplings obey
$2g^*_{R_{BB}}<g^*_{R_{AB}}<2g^*_{R_{AA}}$ (when $\delta<1$) it is
appropriate to use the mean field solutions derived in section 2.
For the case where $n_A\gg n_B$, this gives:
\begin{equation}
\langle a\rangle\sim\left({\Gamma(\epsilon/2)\over
(8\pi)^{d/2}}\right)(D_At)^{-d/2},
\end{equation}
and
\begin{equation}
\langle b\rangle\sim F(D_At)^{-\beta},
\end{equation}
with
\begin{equation}
\beta\approx {d\over 2}\left({1+\delta \over 2}\right)^{d/2} \qquad
F\approx n_B\left({\Gamma(\epsilon/2)\over
n_A(8\pi)^{d/2}}\right)^{\left({
1+\delta\over 2}\right)^{d/2}},
\end{equation}
valid for $n_A^{-2/d}D_A^{-1}\ll t\ll t_{1}$, where
\begin{equation}
D_At_{1}\approx\left({n_B\over n_A^{\left({1+\delta
\over
2}\right)^{d/2}}}\right)^{{2\over d}\left(\left({1+\delta\over
2}\right)^{d/2}-1\right)^{-1}}.
\end{equation}
These modified crossover times are obtained by using the expressions
for the running couplings/densities in the mean field
crossovers. Notice that the density decay exponents derived here are
the same as those obtained from the
Smoluchowski approach. However, as we are performing an $\epsilon$
expansion, we are only strictly justified in retaining leading order
$\epsilon$ terms. Consequently we find, for the minority species
density decay exponent and amplitude:
\begin{equation}
\beta=\left({1+\delta\over 2}\right)+O(\epsilon) \qquad
F=n_B\left({1\over 4\pi\epsilon
n_A}+O(\epsilon^0)\right)^{\left({1+\delta
\over 2}\right)+O(\epsilon)}.
\end{equation}
Eventually, however, as the A particles are decaying away more quickly
than the B particles (due to their greater diffusivity when
$\delta<1$), we crossover to a
second regime where $\langle b\rangle\gg\langle a\rangle$. For
$0<\delta<1$, we have:
\begin{equation}
\langle b\rangle\sim\left({\Gamma(\epsilon/2)\over
(8\pi)^{d/2}}\right)
(D_Bt)^{-d/2}
\end{equation}
\begin{equation}
\langle a\rangle\sim E(D_Bt)^{-\alpha},
\end{equation}
with
\begin{eqnarray}
& & \alpha\approx {d\over 2}\left({1+\delta^{-1}\over
2}\right)^{d/2}=
\left({1+\delta^{-1}\over 2}\right)+O(\epsilon) \\ & & E\approx n_A
f(d)\left({\Gamma(\epsilon/2)\over
n_B(8\pi)^{d/2}}\right)^{\left({1+\delta^{-1}\over 2}\right)^{d/2}}
=n_A f(2)\left({1\over
4\pi\epsilon n_B}+O(\epsilon^0)\right)^{\left({1+\delta^{-1}\over
2}\right) +O(\epsilon)}
\end{eqnarray}
where
\begin{equation}
f(d)=\left(1+{[((1+\delta)/2)^{d/2}-1]n_A\over
[\delta^{d/2}-((1+\delta)/2)^{d/2}]n_B}\right)^{-1-\left({1+
\delta^{-1}\over 2}\right)^{d/2}\left({((1+\delta)/2)^{d/2}-
\delta^{d/2}\over ((1+\delta)/2)^{d/2}-1}\right)}.
\end{equation}
This result is valid for $t\gg t_{2}$, where
\begin{equation}
D_Bt_{2}\approx\left({n_Af(d)(1+\delta^{-1})^{d/2}\over
n_B^{((1+\delta^{-1})/2)^{d/2}}}\right)^{{2\over
d}\left(\left({1+\delta^{-1}\over 2}\right)^{d/2}-1\right)^{-1}}.
\end{equation}

Note that for $\delta=1$ the first crossover time
$t_{1}\rightarrow\infty$ - in this case the two species
decay away at the same rate, and so no further crossover
occurs. Alternatively if $\delta=0$, then the first regime
is left, but the second crossover time $t_{2}\rightarrow\infty$. In
that case the minority species finally decays away in the
exponential fashion predicted in \cite{PG}. For the intermediate case
where $\delta$ is small, but nonzero, the decay exponent for the
minority species becomes large in the final regime. The
explanation for this result lies in the relatively large diffusivity
of the minority A species (if $D_A$ is large) and/or the increased
density amplitude for the majority B particles (if $D_B$ is
small). Both these effects will lead to an increased rate of decay for
the A species.

Finally, if the initial conditions are changed such that now $n_B\gg
n_A$, with $\delta\neq 0$, then we obtain the same results as for
the second of the above regimes for $D_Bt\gg n_B^{-2/d}$,
with $f\approx 1$.

\subsection{One Loop Results}

We now describe the one loop improvements to the tree level result. In
the regime $\langle a\rangle\gg\langle b\rangle$, the dominant
diagrams will be those where the
minimum possible number of $\langle b\rangle_{cl}$ insertions are
made. For the majority A species the appropriate diagram is shown in
figure 5, where there are no $\langle b\rangle_{cl}$ insertions. This
is identical to the one loop diagram for
$A+A\rightarrow\emptyset$ evaluated in \cite{L}, which gives, in
conjunction with the subleading terms from the tree level:
\begin{equation}
\langle a\rangle\sim\left({1\over 4\pi\epsilon}+{2\ln 8\pi-5\over
16\pi}+O(\epsilon)\right) (D_At)^{-d/2}. \label{maxexp}
\end{equation}
In addition, for that subset of diagrams with no $\langle
b\rangle_{cl}$ insertions, the decay exponent is exact.
More details of this calculation, including a demonstration of the
cancellation of divergences, can be found in \cite{L}.

Turning now to the one loop calculation for the minority species, the
appropriate diagrams are the three shown in figure 6, each of which
contains just one $\langle b\rangle_{cl}$ insertion:
\begin{eqnarray}
(i) & & \qquad {-4\lambda_{AB}\lambda_{AA}^2n_A^2n_B\over
(2\lambda_{AA}n_At)^{1+\lambda_{AB}/2\lambda_{AA}}}\int{d^dk\over
(2\pi)^d} \int_0^t
dt_2\int_0^{t_2} dt_1 (t-t_2) \nonumber
\\ & & \qquad\qquad\qquad\qquad\qquad
\times{(1+2\lambda_{AA}n_At_1)^2\over
(1+2\lambda_{AA}n_At_2)^3}\exp{[-2k^2(t_2-t_1)]} \label{la1}\\
(ii) & & \qquad {-2\lambda_{AB}^2\lambda_{AA}n_A^2n_B\over
(2\lambda_{AA}n_At)^{\lambda_{AB}/2\lambda_{AA}}}\int{d^dk\over
(2\pi)^d} \int_0^t dt_2\int_0^{t_2} dt_1\int_{t_1}^{t_2}
dt'{(1+2\lambda_{AA}n_At_1)^2\over (1+2\lambda_{AA}n_At_2)^2}
\nonumber \\ & & \qquad\qquad\times{1\over
(1+2\lambda_{AA}n_At')^2}
\exp{[-k^2(t_2(1+\delta)-2t_1+(1-\delta)t')]} \label{la2} \\
(iii) & & \qquad {\lambda_{AB}^2n_An_B\over
(2\lambda_{AA}n_At)^{\lambda_{AB}/2\lambda_{AA}}}\int{d^dk\over
(2\pi)^d}\int_0^t dt_2\int_0^{t_2} dt_1 {(1+2\lambda_{AA}n_At_1)\over
(1+2\lambda_{AA}n_At_2)^2} \nonumber \\ & & \qquad\qquad\qquad\qquad
\qquad\qquad\qquad \times\exp{[-k^2(1+\delta)(t_2-t_1)]}. \label{la3}
\end{eqnarray}

The detail of the evaluation of these diagrams is rather
subtle. Essentially we are interested in extracting the most divergent
parts of these integrals, which will turn out to be pieces of
$O(\epsilon^{-1})$ and $O(\epsilon^0)$. However, we must be careful
not to confuse genuine bare divergences (of $O(\epsilon^{-1})$ which
must be removed by the renormalisation of the theory), with
logarithmic pieces, which we must retain. The divergences arise in
diagrams (i) and (iii) as the difference in time $t_2-t_1$ between
the beginning and end of the loops tends to zero (in
$d=2$). After the process of renormalisation we find corrections of
the form:
\begin{equation}
1+({\rm constant})\epsilon\ln(({\rm constant})t^{d/2})+O(\epsilon^2).
\end{equation}
If this series is identified as the expansion of an exponential, then
we find that our one loop diagrams
(together with subleading components from the tree level) have
provided $O(\epsilon)$ corrections to the exponents.

Diagrams (i) and (iii) are relatively straightforward to evaluate. The
$k$ and $t_1$ integrals are elementary, and the final $t_2$ integrals
can be done by parts to extract the necessary most divergent pieces
(up
to $O(\epsilon^0)$). The second diagram of figure 6 is more
complicated, and we perform its evaluation in appendix B - although we
are only able to extract the logarithmic piece of $O(
t^{-\lambda_{AB}/2\lambda_{AA}}\,t^{\epsilon/2}\ln t)$.
There will be corrections to this of $O(
t^{-\lambda_{AB}/2\lambda_{AA}}\,t^{\epsilon/2})$
(contributing to a modified amplitude) which we have been
unable to calculate. We find asymptotically:
$$
\;(i)\;{-\lambda_{AB}n_B\over
8\pi(2\lambda_{AA}n_At)^{\lambda_{AB}/2\lambda_{AA}}}\left({2t^{
\epsilon/2}(\ln(2\lambda_{AA}n_At)-1)\over\epsilon}+t^{\epsilon/2}
(\ln(2\lambda_{AA}n_At)-1)\ln(8\pi)\right.
$$
\begin{equation}
\left.\quad +{15t^{\epsilon/2}\over 4}-{3\over 2}t^{\epsilon/2}
\ln(2\lambda_{AA}n_At)-\int_0^t t_2^{-1+\epsilon/2}\ln(1+
2\lambda_{AA}n_At_2)dt_2+O(\epsilon)\right)\label{q1}
\end{equation}
\begin{equation}
\;(ii)\;{-\lambda_{AB}^2n_B\over 32\pi\lambda_{AA}(2\lambda_{AA}n_At)
^{\lambda_{AB}
/2\lambda_{AA}}}\left(\delta+{1\over 2}(\delta^2-1)\left[\ln\left(
{1-\delta\over1+\delta}\right)\qquad\qquad\qquad\qquad\quad
\right.\right.
\end{equation}
$$
\left.\left. \quad\qquad\qquad\qquad\qquad\qquad
-\int_{-1}^{1-\delta\over
1+\delta} dv {(1+v)^2\over v^2}\ln(1+v)\right]+O(\epsilon)\right)
t^{\epsilon/2}\ln(2\lambda_{AA}n_At)
$$
\begin{equation}
\;(iii)\;{\lambda_{AB}^2n_B (4\pi(1+\delta))^{-1}\over
2\lambda_{AA}(2\lambda_{AA}n_At)^{\lambda_{AB}/2\lambda_{AA}}}
\left({2t^{\epsilon/2}\ln(2\lambda_{AA}n_At)\over
\epsilon}+t^{\epsilon/2}\ln(2\lambda_{AA}n_At)
\ln(4\pi(1+\delta))\right.\label{q2}
\end{equation}
$$
\left.\quad
-t^{\epsilon/2}(\ln(2\lambda_{AA}n_At)-1)-\int_0^t
t_2^{-1+\epsilon/2}\ln(1+2\lambda_{AA}n_At_2)dt_2
+O(\epsilon)\right).
$$
To one loop accuracy we can make the replacement:
$\lambda_{\{ij\}}=\kappa^{\epsilon}g_{0_{\{ij\}}}\rightarrow
\kappa^{\epsilon} g_{R_{\{ij\}}}$.
These results must now be combined with the subleading
terms from the tree level. Using (\ref{gexp}), we find
\begin{eqnarray}
& &\langle b\rangle\sim {n_B\over
(2\lambda_{AA}n_At)^{\lambda_{AB}/2\lambda_{AA}}}={n_B\over
(2\kappa^{\epsilon}g_{R_{AA}}n_At)^{g_{R_{AB}}/2g_{R_{AA}}}}\left(1
-{g_{R_{AB}}\over 2g^*_{R_{AA}}} \right.\nonumber \\ & & \left.
\quad -{g^2_{R_{AB}}\over
2g_{R_{AA}}g^*_{R_{AB}}}\ln(2\kappa^{\epsilon}g_{R_{AA}}n_At)+
{g_{R_{AB}}\over
2g^*_{R_{AA}}}\ln(2\kappa^{\epsilon}g_{R_{AA}}n_At)+O(g_R^2)\right).
\label{q3}
\end{eqnarray}
If we now insert explicit $\epsilon$ expanded values for the fixed
points $g^*_{R_{\{ij\}}}$,
then we discover that the bare divergences cancel between (\ref{q1}),
(\ref{q2}), and (\ref{q3}). With insertion into
the Callan-Symanzik solution (\ref{CSS}), we also find that the pieces
we have left as integrals in (i) and (iii) (which are $O(
t^{\epsilon/2}(\ln t)^2)$) also mutually cancel. Eventually we find:
$$
\langle b\rangle\sim({\rm const.}) t^{-{d\over
2}\left({1+\delta\over
2}\right)^{d/2}}\left(1+{\epsilon(1+\delta)\over
8}\left[1-2(1+\delta)\left({\delta\over
4}+{\delta^2-1\over 8}\left(\ln\left({1-\delta\over1+\delta}\right)
\right.\right.\right.\right.
$$
\begin{equation}
\left.\left.\left.\left.\qquad\qquad
\qquad\qquad -\int_{-1}^{1-\delta\over
1+\delta}dv{(1+v)^2\over v^2}\ln(1+v)\right)
\right)\right]\ln(({\rm const.})t^{d/2})+O(\epsilon^2)\right),
\label{bfirst}
\end{equation}
where we have neglected $O(\epsilon)$ pieces which, aside from the
prefactor, are time {\it independent}. These terms
contribute only to the density amplitude. We now evaluate the integral
in (\ref{bfirst}), using
$$
\int_{-1}^{1-\delta\over 1+\delta}{\ln(1+v)\over v}dv=\int_0^{2
\over 1+\delta}{\ln u\over u-1}du=\int_0^1{\ln u\over u-1}du+\int_1^
{2\over 1+\delta}{\ln u\over u-1}du
$$
\begin{equation}
\qquad\qquad\qquad\qquad\qquad ={\pi^2\over 6}-f\left\{{2\over
1+\delta}\right\}, \label{minexp}
\end{equation}
where $f\{x\}$ is the dilogarithm function \cite{AS}. The other parts
of
the integral are elementary. The next step is to $\epsilon$ expand the
RG improved tree level result:
\begin{equation}
{d\over 2}\left({1+\delta\over 2}\right)^{d/2}=\left({1+\delta\over
2}\right)\left(1-
{\epsilon\over 2}\left(1+\ln\left({1+\delta\over 2}\right)\right)+
O(\epsilon^2)\right). \label{tei}
\end{equation}
Then, exponentiating the $\epsilon$ expansion in (\ref{bfirst}), we
find $\langle b\rangle=O(t^{-\beta})$, where
\begin{eqnarray}
& & \beta=\left({1+\delta\over 2}\right)\left(1-{\epsilon\over
2}\left[{3\over 2}+\ln\left({1+\delta\over 2}\right)-
{\delta(1+\delta)\over 4}\left[1+2\ln\left({1+\delta\over
2}\right)\right]
\right.\right. \nonumber \\ & & \left.\left.\qquad\qquad
-{1\over 4}(\delta^2-1)\left(1+(1+\delta)\left[f\left\{{2\over
1+\delta}\right\}-{\pi^2\over 6}\right]\right)\right]\right)
+O(\epsilon^2). \label{bsecond}
\end{eqnarray}
$\beta$ is plotted as a function of $\delta$ for $\epsilon=1$ ($d=1$)
in figure 9. For the case where $\delta=1$, we recover the decay rate
$\langle b\rangle=O(t^{-d/2})$. This is to be expected, as when
$\delta=1$ we are effectively again dealing with a single species
reaction-diffusion system (at least for $d<2$). In that case the
density decay
exponent is known to all orders in perturbation theory \cite{L}, and
is
in agreement with our result. For the case where $\delta=0$ and $d=1$,
the decay exponent is also known exactly to be $\langle b\rangle=O(
t^{-0.375})$ \cite{D3}. This can be compared with our result, where we
find
\begin{equation}
\beta={1\over 16}+{1\over 4}\ln 2 +{\pi^2\over 64}\approx 0.39
\qquad (\delta=0).
\end{equation}
Consequently, this answer is a modest improvement over the
Smoluchowski result derived in section 2, and also in \cite{KBR}.

For the case $n_B\gg n_A$ (and hence $\langle b\rangle\gg\langle
a\rangle$), we could follow
the same route as described above, by evaluating the one loop diagrams
shown in figures 7 and 8. However, as we mentioned in the last section
we can much more easily
obtain these corrections by swapping the labels on the A
and B particles, and then relabeling:
$$
n_A\leftrightarrow n_B \quad \lambda_{AA}\leftrightarrow\lambda_{BB}
\quad D_A\leftrightarrow D_B.
$$
Following this procedure, the majority species amplitude/exponent can
be found by taking $D_A\rightarrow D_B$ in equation (\ref{maxexp}):
\begin{equation}
\langle b\rangle\sim\left({1\over 4\pi\epsilon}+{2\ln 8\pi-5\over
16\pi}+O(\epsilon)\right)(D_Bt)^{-d/2}. \label{ola}
\end{equation}
We can obtain the one loop minority species
exponent by substituting $\delta\rightarrow \delta^{-1}$ in
equation (\ref{bsecond}):
\begin{equation}
\langle a\rangle=O(t^{-\alpha}),
\end{equation}
where
$$
\alpha=\left({1+\delta^{-1}\over 2}\right)\left(1-{\epsilon\over
2}\left[{3\over 2}+\ln\left({1+\delta^{-1}\over 2}\right)-
{\delta^{-1}(1+\delta^{-1})\over
4}\left[1+2\ln\left({1+\delta^{-1}\over 2}\right)\right]
\right.\right.
$$
\begin{equation}
\left.\left. -{1\over
4}(\delta^{-2}-1)\left(1+(1+\delta^{-1})\left[f\left\{{2\over
1+\delta^{-1}}\right\}-{\pi^2\over 6}\right]\right)\right]\right)
+O(\epsilon^2). \label{olalph}
\end{equation}
Notice, however, that in forming the one loop corrections for the
minority species exponent, we have had to expand the RG improved tree
level result:
\begin{equation}
{d\over 2}\left({1+\delta^{-1}\over 2}\right)^{d/2}=
\left({1+\delta^{-1}\over 2}\right)\left(1-{\epsilon\over
2}\left(1+\ln\left({1+\delta^{-1}\over 2}\right)\right)+O(\epsilon^2)
\right). \label{treexp1}
\end{equation}
The error arising from this expansion will become large as $\delta$
becomes small. Eventually this inaccuracy will cause the exponent to
reach a maximum and then {\it decrease} as $\delta$ is further reduced
- behaviour which is clearly
unphysical. In order to reduce the error, and to ensure that the
expansion in equation (\ref{treexp1}) is qualitatively correct, we
need to retain the $O(\epsilon^2)$ terms. Hence the
one loop exponent in equation (\ref{olalph}) should be treated
with some caution - terms of order $O(\epsilon^2)$ will probably be
required for precise results. Consequently the (non $\epsilon$
expanded) RG
improved tree level result given in the last section may be
more accurate in this regime. In figure 10 we have
plotted the one loop exponent $\alpha$ as a function of $\delta$, for
$d=1$ ($\epsilon=1$), in the region $0.7\leq\delta\leq
1$, where the exponent is still {\it increasing} for decreasing
$\delta$.

In principle, calculations can also be made for the case
with $n_A\gg n_B$, but where we have crossed over to the
regime $\langle b\rangle\gg\langle a\rangle$ (for
$0<\delta<1$ and times $t\gg t_{2}$). However, a rigorous evaluation
of the one loop diagrams
is now much more difficult, as the functional forms for the densities
and response functions will change over time. Nevertheless, since the
above corrections to the exponents come from asymptotic logarithmic
terms, it is plausible to suppose that the new exponent
corrections will be dominated by contributions from the final
asymptotic regime. If this is indeed
the case, then the one loop exponents (though {\it not} the
amplitudes) will be unchanged from the previous results (equations
(\ref{ola}) to (\ref{olalph})). This calculation will, however, suffer
from the same problem as described above.

\subsection{$d=d_c$}

For the case $d=d_c=2$ we expect logarithmic corrections to the decay
exponents, as the reaction rates $\lambda_{\{ij\}}$ are marginal
parameters at the critical dimension. We can find the running
couplings from the characteristic equation (\ref{gchar}) by taking the
limit $\epsilon\rightarrow 0$ in equations
(\ref{b1}), (\ref{b2}), and (\ref{b3}):
\begin{equation}
\tilde g_{R_{AA}}(\kappa^{-2})={g_{R_{AA}}\over
1+g_{R_{AA}}C\ln(\kappa^2t)}\sim (C\ln t)^{-1}
\end{equation}
\begin{equation}
\tilde g_{R_{BB}}(\kappa^{-2})={g_{R_{BB}}\over
1+g_{R_{BB}}C\delta^{-1} \ln (\kappa^2t)}\sim
(C\delta^{-1}\ln t)^{-1}
\end{equation}
\begin{equation}
\tilde g_{R_{AB}}(\kappa^{-2})={g_{R_{AB}}\over
1+g_{R_{AB}}C(1+\delta)^{-1}\ln(\kappa^2t)}\sim
(C(1+\delta)^{-1}\ln t)^{-1},
\end{equation}
where we have taken the asymptotic limits. Corrections to the
asymptotic
running couplings will be an order $(\ln t)^{-1}$ smaller, and
consequently these asymptotic expressions will only be correct at very
large times. Hence our expressions for the densities will only be
valid
when both this condition, and the crossover time constraints given
below, are satisfied. In what follows we shall assume the validity of
the first of these two conditions. Notice that the asymptotic running
couplings are still ordered $2\tilde g_{R_{BB}}<\tilde
g_{R_{AB}}<2\tilde
g_{R_{AA}}$ for $\delta<1$, so we can use the mean field solutions
derived in section 2 as the basis for the RG improved tree level
exponents and
amplitudes. Making use of the Callan-Symanzik solution (\ref{CSS}) and
the above running couplings, we find for $\langle a\rangle\gg\langle
b\rangle$:
\begin{eqnarray}
& & \langle a\rangle\sim {\ln t\over 8\pi D_At} \\
& & \langle b\rangle\sim {n_B\over (8\pi
n_AGD_At/\ln t)^{(1+\delta)/2}},
\end{eqnarray}
where $G=\exp\left({4\pi\over
g_{R_{AA}}}\left(1-{(1+\delta)g_{R_{AA}}\over
g_{R_{AB}}}\right)\right)$ is a non-universal amplitude
correction. Note that the
next order terms for the minority species are suppressed by a factor
of only $(\ln \ln t)/(\ln
t)$. Using our expressions for the running couplings/densities in the
mean field crossovers, we find that these expressions are valid for
times
$D_A^{-1}n_A^{-1}\ln t\ll t\ll T_1$, where
\begin{equation}
(D_AT_1/\ln T_1)\approx\left({(Gn_A)^{(1+\delta)/2}\over
n_B}\right)^{2\over 1-\delta}.
\end{equation}
For the case $\delta<1$ the system will eventually enter a second
regime, where now the B species will be in the majority. We have (for
$\delta\neq 0$):
\begin{eqnarray}
& & \langle b\rangle\sim{\ln t\over 8\pi D_Bt} \\
& & \langle a\rangle\sim{n_A K\over (8\pi n_BHD_Bt/\ln
t)^{(1+\delta^{-1})/2}},
\end{eqnarray}
with
\begin{equation}
H=\exp\left({4\pi\delta\over
g_{R_{BB}}}\left(1-{(1+\delta^{-1})g_{R_{BB}}\over
g_{R_{AB}}}\right)\right) \qquad
K=\left(1+{n_A\over n_B}\right)^{\delta^{-1}-1\over 2}.
\end{equation}
This is valid for times when $t\gg T_2$, where
\begin{equation}
(D_BT_2/\ln T_2)\approx\left({(Hn_B)^{(1+\delta^{-1})/2}\over
n_A(1+\delta^{-1})K}\right)^{2\over 1-\delta^{-1}}.
\end{equation}
Alternatively, if we begin with $n_B\gg n_A$, then for $\delta\neq 0$
and $(D_Bt/\ln t)\gg n_B^{-1}$,
we have the same results as for the second of the above cases,
with $K\approx 1$. Interestingly, the logarithmic corrections we have
derived in this section using the RG approach differ
slightly from the Smoluchowski results given in section 2.

\section{Conclusion}

In this paper we have made a comparison of two methods for treating
fluctuation effects in a reaction-diffusion system. We have found that
the Smoluchowski and field theory approaches are rather similar - the
Smoluchowski approximation, for $d<2$, giving the same exponents as
the renormalisation group improved tree level in the field theory. In
addition, we have gone on to calculate the field theoretic one loop
corrections, which have yielded improved values for the exponents. The
advantage of the field theory is that it provides a systematic way to
calculate these corrections - a procedure which is lacking in the
Smoluchowski approach. Furthermore the use of renormalisation group
techniques has demonstrated universality in the asymptotic amplitudes
and exponents, in that, for $d<2$, they only depend on the
diffusivities and the initial densities, and not on the reaction
rates.

The theory we have developed in this paper can easily be extended to
slightly different situations. Consider first an
annihilation/coagulation reaction-diffusion system, where the
following reactions occur:
$$
A+A\rightarrow A\qquad B+B\rightarrow B\qquad A+B\rightarrow\emptyset.
$$
The Smoluchowski approach differs from before only in the absence of
factors of $2$ in the rate equation terms describing the same species
reactions. Consequently, if we begin with $n_A\gg n_B$ then the
minority species will decay as
\begin{equation}
b=O\left(t^{-d\left({1+\delta\over 2}\right)^{d/2}}\right) \qquad
(d<2).
\end{equation}
On the other hand, the field theory description lacks only the
factors of $2$ in the action (\ref{action}). If this
difference is followed through
then the decay exponent in the RG improved tree level is seen to be
the same as in the
Smoluchowski approach. However, this difference of a factor of $2$ has
a major effect on the response functions (where this factor appears as
a power), and as a result the new one loop corrections will
be different from those calculated in section 4.2. These results
should be compared with the exact solution \cite{F1,F2,BAV} for the
minority species decay rate $b=O(t^{-\gamma})$, where:
\begin{equation}
\gamma={\pi\over 2\cos^{-1}(\delta/(1+\delta))}.
\end{equation}
Note that in this case, although the Smoluchowski answer is
qualitatively correct, it deviates considerably from the exact answer.
Hence we can see that application of the Smoluchowski approach
does not always lead to accurate exponents.

Another possible extension is to consider reaction-diffusion systems
with more than two species of particle. For example, examining a three
species system, we could have the reactions:
$$
A+A\rightarrow\emptyset \qquad A+B\rightarrow\emptyset \qquad
A+C\rightarrow\emptyset
$$
$$
B+B\rightarrow\emptyset \qquad B+C\rightarrow\emptyset \qquad
C+C\rightarrow\emptyset.
$$
Analysis of this situation is very similar to before, and we merely
remark that in the appropriate asymptotic regimes the Smoluchowski and
RG improved tree level exponents
(consisting of ratios of diffusion constants) are once again
identical. Hence the convergence between the
Smoluchowski exponents and those obtained from the RG improved tree
level is fairly robust, and is not simply confined to the two
species systems we have previously been considering. A further
possibility is to analyse the case where we have a continuous
distribution of diffusivities, but with only a {\it
single} reaction channel. This has been studied from the Smoluchowski
point of view by Krapivsky {\it et al.} \cite{KBR}, and it would be
interesting to extend our RG methods to include this situation.

Our theory could also be employed to consider
clustered immobile reactants - a generalisation of the
$\delta=0$ case included in our calculations. This situation has been
analysed by Ben-Naim \cite{B}, using the
Smoluchowski approach, where the dimension of the cluster $d_I$ was
found to substantially affect the kinetics. Specifically, for
codimensionality $d-d_I<2$ (in a space of dimension $d$) a finite
fraction of the impurities was found to survive,
whereas for $d-d_I\geq 2$ the clusters decayed away indefinitely. The
formalism we have presented in this paper could be adapted to
study this clustered impurity problem,
where calculations could be made without reliance on
the Smoluchowski approach.

\

\noindent{\bf Acknowledgments.}
\noindent  The author thanks John Cardy for suggesting this problem
and for many useful discussions. Financial support from the EPSRC is
also acknowledged.

\section{Appendix A: Response Functions}
Obtaining an exact analytic expression for the response functions is,
in general, very hard. Suppose we define the ``trunk'' to be the line
of
propagators onto which the density lines are attached, as shown at the
bottom of figure
4. Difficulties arise from diagrams where the ``trunk'' changes
from one propagator into the other, and then back again, as shown in
last of the diagrams for the $L$ response function in figure 4. If
diagrams of this type are initially
excluded then progress can be made. Consider first the two
subseries shown in figure 11, for the functions $\xi(k,t_2,t_1)$ and
$\theta(k,t_2,t_1)$, where diagrams of the above kind have been
excluded. These series can be summed exactly (using
the same technique as described in \cite{L}), giving:
\begin{equation}
\xi(k,t_2,t_1)=\exp{\left(-k^2(t_2-t_1)\right)}\exp{\left(-\int_{t_1}^
{t_2}(4\lambda_{AA}a+\lambda_{AB}b)dt\right)}
\end{equation}
\begin{equation}
\theta(k,t_2,t_1)=\exp{\left(-k^2(t_2-t_1)\delta\right)}\exp{\left(
-\int_{t_1}^{t_2}(4\lambda_{BB}b+\lambda_{AB}a)dt\right)}.
\end{equation}
The full response functions are now given by the diagrammatic
equations shown in figure 12, where all possible diagrams are
included. Written out explicitly these give:
\begin{eqnarray}
& & L(k,t_2,t_1)=\xi(k,t_2,t_1)-\lambda_{AB}\int_{t_1}^{t_2}
\xi(k,t_2,\tau) a(\tau) M(k,\tau,t_1)d\tau \\
& & M(k,t_2,t_1)=-\lambda_{AB}\int_{t_1}^{t_2}\theta(k,t_2,\tau)
b(\tau)L(k,\tau,t_1)d\tau \\
& & N(k,t_2,t_1)=\theta(k,t_2,t_1)-\lambda_{AB}\int_{t_1}^{t_2}
\theta(k,t_2,\tau) b(\tau) P(k,\tau,t_1)d\tau \\
& & P(k,t_2,t_1)=-\lambda_{AB}\int_{t_1}^{t_2}\xi(k,t_2,\tau)
a(\tau)N(k,\tau,t_1)d\tau.
\end{eqnarray}
In general this set of coupled integral equations is intractable -
however we can make progress in the limit where $\langle a\rangle\gg
\langle b\rangle$, or $\langle b\rangle\gg \langle
a\rangle$. Considering the case where $\langle a\rangle\gg
\langle b\rangle$, the dominant
contributions to the response functions come from diagrams with
the minimum possible number of $\langle b\rangle_{cl}$ density line
insertions. Accordingly, we can now truncate the full diagrammatic
equations, as shown in figure 13. Notice that to this order $L$, $N$,
and $P$ contain no $\langle b\rangle_{cl}$ density insertions, whereas
$M$ must contain
one such insertion. In this approximation we can now perform the
integrals inside the $\xi$ and $\theta$ functions, using the
appropriate mean field density:
\begin{equation}
\int_{t_1}^{t_2}(4\lambda_{AA}a+\lambda_{AB}b)dt\approx\int_{t_1}
^{t_2}{4\lambda_{AA}n_A\over 1+2\lambda_{AA}n_At}dt=\ln\left({
1+2\lambda_{AA}n_At_2\over 1+2\lambda_{AA}n_At_1}\right)^2
\end{equation}
\begin{equation}
\int_{t_1}^{t_2}(4\lambda_{BB}b+\lambda_{AB}a)dt\approx\int_{t_1}
^{t_2}{\lambda_{AB}n_A\over 1+2\lambda_{AA}n_At}dt=\ln\left({
1+2\lambda_{AA}n_At_2\over 1+2\lambda_{AA}n_At_1}\right)^{
\lambda_{AB}/2\lambda_{AA}},
\end{equation}
and therefore
\begin{equation}
\xi(k,t_2,t_1)=\left({1+2\lambda_{AA}n_At_1\over 1+2\lambda_{AA}
n_At_2}\right)^2\exp{(-k^2(t_2-t_1))}
\end{equation}
\begin{equation}
\theta(k,t_2,t_1)=\left({1+2
\lambda_{AA}n_At_1\over 1+2\lambda_{AA}n_At_2}\right)^{\lambda_{AB}/
2\lambda_{AA}}\exp{(-k^2(t_2-t_1)\delta)}.
\end{equation}
Using these expressions, it is now straightforward to derive the
response functions given in equations (\ref{R1L}), (\ref{R1N}),
(\ref{R1P}), and (\ref{R1M}).

\section{Appendix B: A One Loop Integral}

For the case where $\langle a\rangle\gg\langle b\rangle$ the hardest
of the three diagrams of
figure 6 to evaluate is (ii) - see equation (\ref{la2}). We shall
evaluate it first in $d=2$, and then deduce its form in
$d=2-\epsilon$. Notice that the extra integration resulting from the
$\langle b\rangle_{cl}$ insertion in the loop ensures that this
diagram is not divergent. Taking the asymptotic part of the $t_1$ and
$t'$ pieces, we find:
$$
{-\lambda_{AB}^2n_B\over
2\lambda_{AA}(2\lambda_{AA}n_At)^{\lambda_{AB}/2\lambda_{AA}}}\int
{d^2k\over (2\pi)^2}\int_0^t dt_2\int_0^{t_2}dt_1\int_{t_1}^{t_2}
dt'{(2\lambda_{AA}n_A)^2 t_1^2\over (1+2\lambda_{AA}n_At_2)^2\,t'^2}
$$
\begin{equation}
\qquad\qquad\qquad\qquad\qquad\qquad\times\exp{(-k^2(t_2(1+\delta)
-2t_1+(1-\delta)t'))}.
\end{equation}
The $k$ and $t'$ integrals are elementary, giving
\begin{equation}
{-\lambda_{AB}^2n_B\over
8\pi\lambda_{AA}(2\lambda_{AA}n_At)^{\lambda_{AB}/2\lambda_{AA}}}
\int_0^t{dt_2 (2\lambda_{AA}n_A)^2\over (1+2\lambda_{AA}n_At_2)^2}
\int_0^{t_2} dt_1 t_1^2
\end{equation}
$$ \times\left({1\over (t_2(1+\delta)-2t_1)}\left[{1\over
t_1}-{1\over t_2}\right]+{1-\delta\over
(t_2(1+\delta)-2t_1)^2}\ln\left({2t_1\over
(1+\delta)t_2}\right)\right).
$$
Although the first part of the $t_1$ integral is straightforward, the
second piece involving the logarithm is more difficult. However, if we
make the transformation
\begin{equation}
v={2t_1\over (1+\delta)t_2}-1,
\end{equation}
we find:
\begin{equation}
\int_0^{t_2}dt_1 {(1-\delta)t_1^2 \over
(t_2(1+\delta)-2t_1)^2}\ln\left({2t_1\over (1+\delta)t_2}\right)=
{1\over 8}(1-\delta^2)t_2\int_{-1}^{1-\delta\over
1+\delta}dv{(1+v)^2\over v^2}\ln(1+v),
\end{equation}
where all time dependency has been removed from the integral
limits. The final $t_2$ integral is then easy to perform, and we end
up with:
\begin{equation}
{-\lambda_{AB}^2n_B\over
32\pi\lambda_{AA}(2\lambda_{AA}n_At)^{\lambda_{AB}
/2\lambda_{AA}}}\left(\delta+{1\over 2}(\delta^2-1)\left[\ln\left(
{1-\delta\over1+\delta}\right)\qquad\qquad\qquad\qquad\qquad\quad
\right.\right.
\end{equation}
$$
\left.\left. \quad\qquad\qquad\qquad\qquad\qquad\qquad\qquad
-\int_{-1}^{1-\delta\over
1+\delta} dv {(1+v)^2\over v^2}\ln(1+v)\right]\right)
\ln(2\lambda_{AA}n_At).
$$
However, we now need to extend this analysis to determine the
behaviour of the integral in $d=2-\epsilon$. If we take the asymptotic
part of all the pieces inside the integral, and perform power
counting,
we find that it should scale as
$t^{-\lambda_{AB}/2\lambda_{AA}}\,t^{\epsilon/2}$. However,
this procedure is not strictly valid, as in moving to the asymptotic
version a false
$t_2=0$ divergence is created. Nevertheless, the integral is dominated
by contributions from late times where arguments based on power
counting should be valid. Hence in $d=2-\epsilon$ we find:
\begin{equation}
{-\lambda_{AB}^2n_B\over
32\pi\lambda_{AA}(2\lambda_{AA}n_At)^{\lambda_{AB}
/2\lambda_{AA}}}\left(\delta+{1\over 2}(\delta^2-1)\left[\ln\left(
{1-\delta\over1+\delta}\right)\qquad\qquad\qquad\qquad\qquad\quad
\right.\right.
\end{equation}
$$
\left.\left. \quad\qquad\qquad\qquad\qquad\qquad
-\int_{-1}^{1-\delta\over 1+\delta} dv {(1+v)^2\over v^2}\ln(1+v)
\right]+O(\epsilon)\right)t^{\epsilon/2}\ln(2\lambda_{AA}n_At).
$$
Further subleading corrections (in time), which we have not
calculated, will lack the logarithm factor, and so
will contribute to the {\it amplitude} for the minority species
density.

\newpage
\listoffigures
\newpage
\begin{figure}
\begin{center}
\leavevmode
\vbox{
\epsfxsize=5in
\epsffile{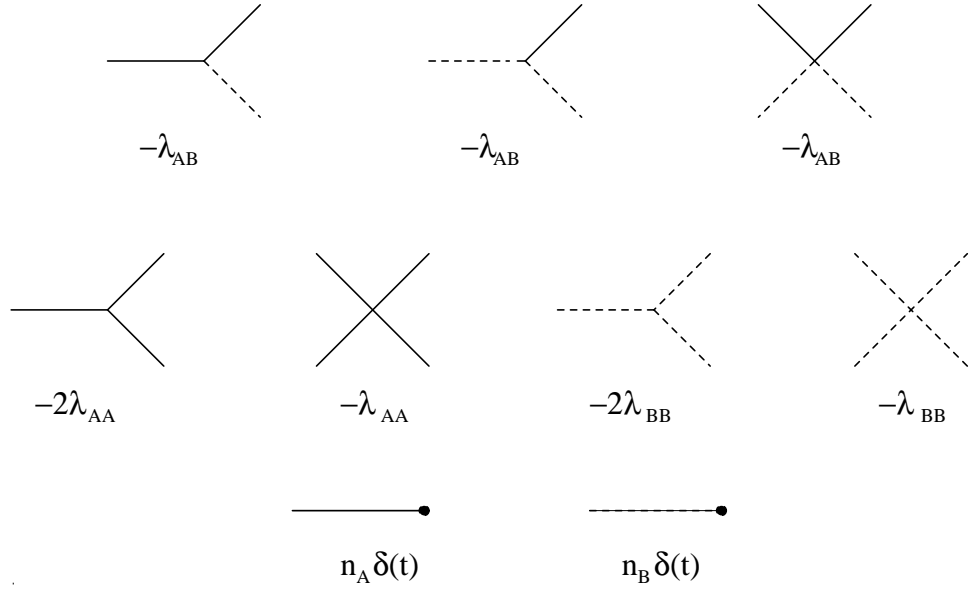}}
\end{center}
\caption{Vertices for the field theory.}
\end{figure}
\begin{figure}
\begin{center}
\leavevmode
\vbox{
\epsfxsize=5in
\epsffile{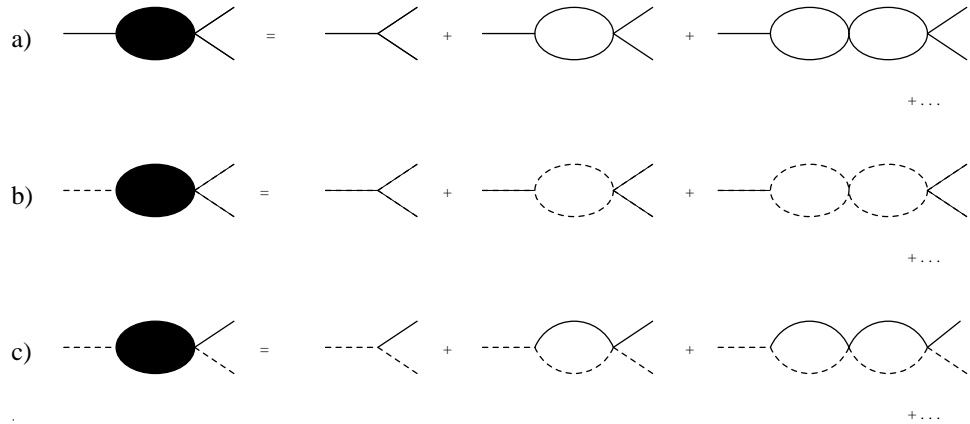}}
\end{center}
\caption{The temporally extended vertex functions (a)
$\lambda_{AA}(k,s)$, (b) $\lambda_{BB}(k,s)$, and (c)
$\lambda_{AB}(k,s)$.}
\end{figure}
\begin{figure}
\begin{center}
\leavevmode
\vbox{
\epsfxsize=5in
\epsffile{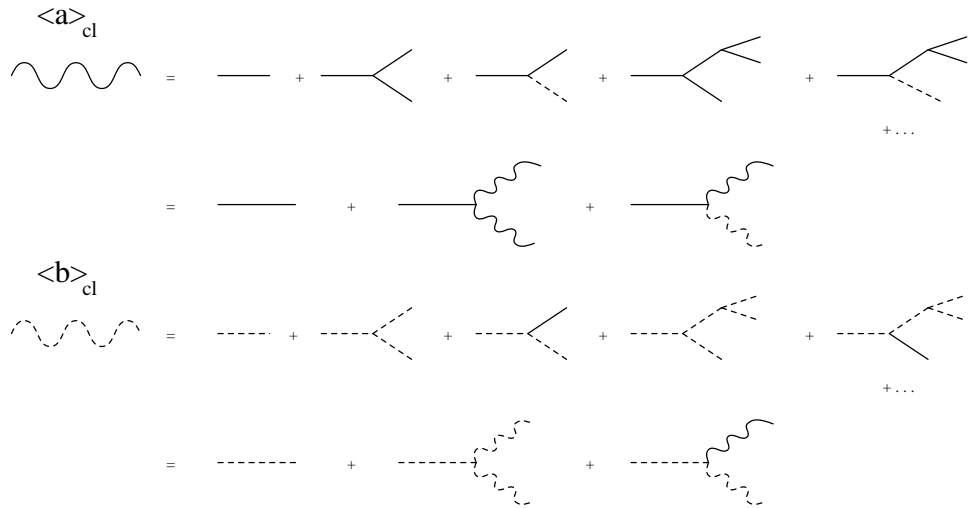}}
\end{center}
\caption{Tree level diagrams for the densities $\langle a\rangle$ and
$\langle b\rangle$.}
\end{figure}
\begin{figure}
\begin{center}
\leavevmode
\vbox{
\epsfxsize=5in
\epsffile{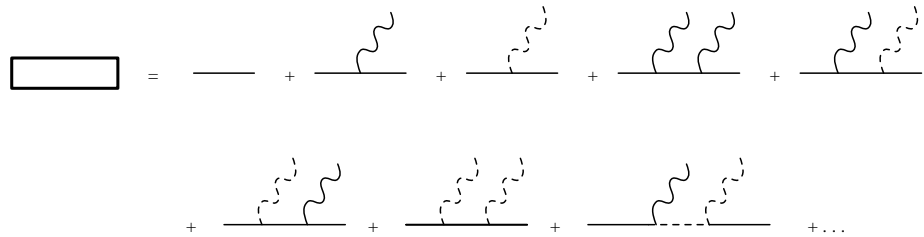}}
\end{center}
\caption{The response functions.}
\end{figure}
\begin{figure}
\begin{center}
\leavevmode
\vbox{
\epsfxsize=2in
\epsffile{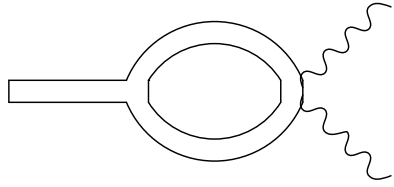}}
\end{center}
\caption{One loop diagram for $\langle a\rangle$ (when $\langle
a\rangle\gg \langle b\rangle$).}
\end{figure}
\begin{figure}
\begin{center}
\leavevmode
\vbox{
\epsfxsize=4in
\epsffile{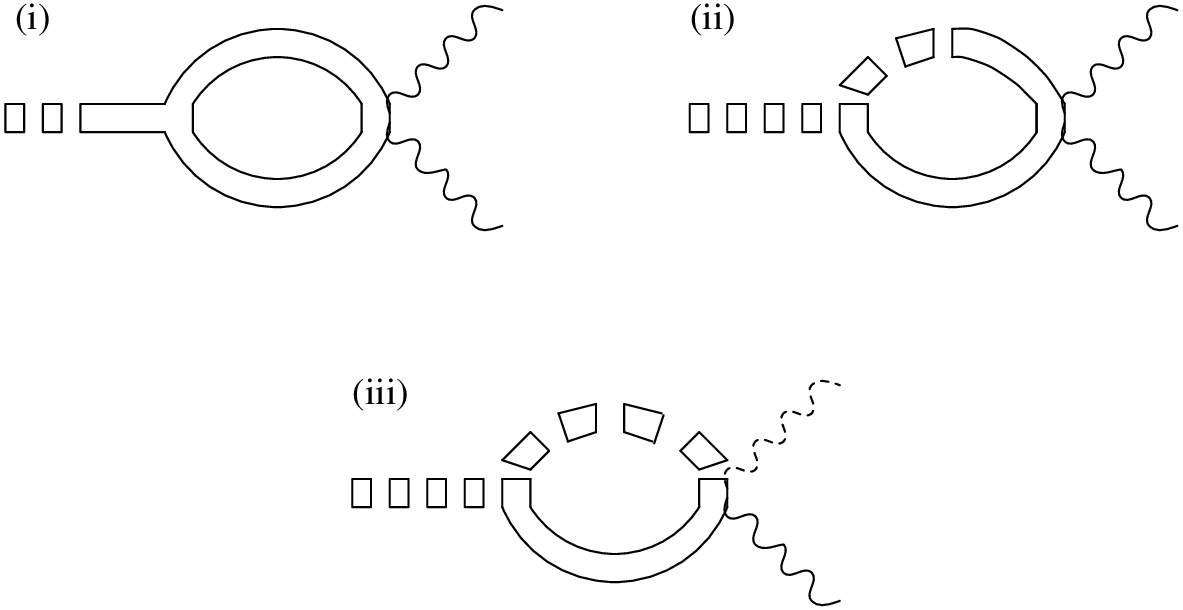}}
\end{center}
\caption{One loop diagrams for $\langle b\rangle$ (when $\langle
a\rangle\gg \langle b\rangle$).}
\end{figure}
\begin{figure}
\begin{center}
\leavevmode
\vbox{
\epsfxsize=3in
\epsffile{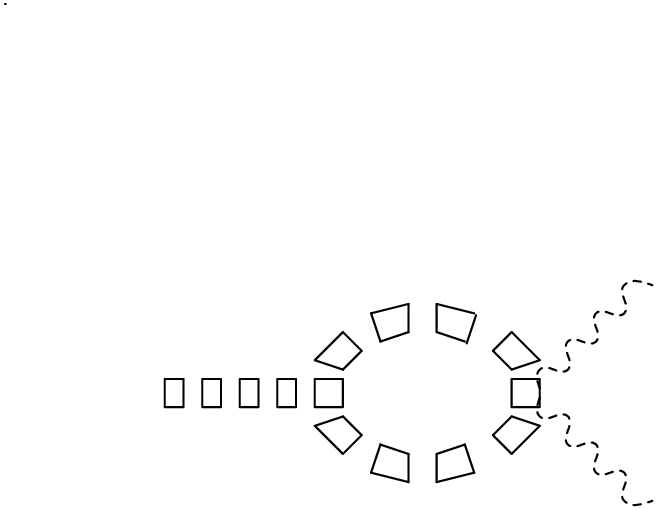}}
\end{center}
\caption{One loop diagram for $\langle b\rangle$ (when $\langle
b\rangle\gg \langle a\rangle$).}
\end{figure}
\begin{figure}
\begin{center}
\leavevmode
\vbox{
\epsfxsize=4in
\epsffile{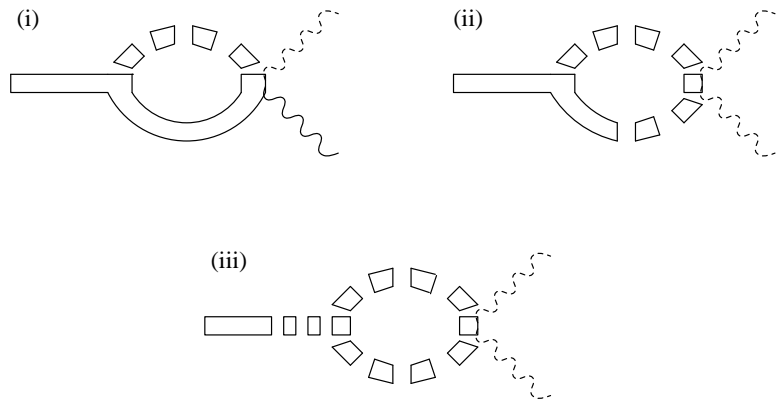}}
\end{center}
\caption{One loop diagrams for $\langle a\rangle$ (when $\langle
b\rangle\gg \langle a\rangle$).}
\end{figure}
\begin{figure}
\begin{center}
\leavevmode
\vbox{
\epsfxsize=5in
\epsffile{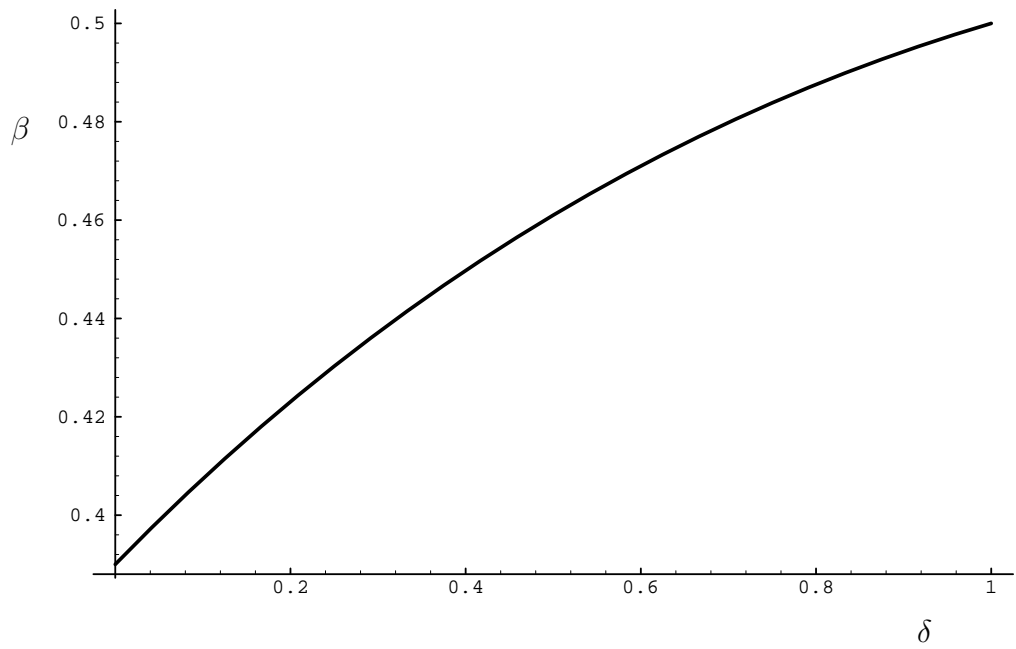}}
\end{center}
\caption{The one loop density decay exponent $\beta$ for the minority
B species ($\langle b\rangle =O(t^{-\beta})$) as a function of
$\delta$ .}
\begin{center}
\vspace{-6.35in}
\
\end{center}
\quad $\beta$
\begin{center}
\vspace{2.15in}
\qquad\qquad\qquad\qquad\qquad\qquad\qquad\qquad\qquad\qquad\qquad
\qquad\quad $\delta$
\end{center}
\begin{center}
\vspace{4.2in}
\
\end{center}
\end{figure}
\begin{figure}
\begin{center}
\leavevmode
\vbox{
\epsfxsize=5in
\epsffile{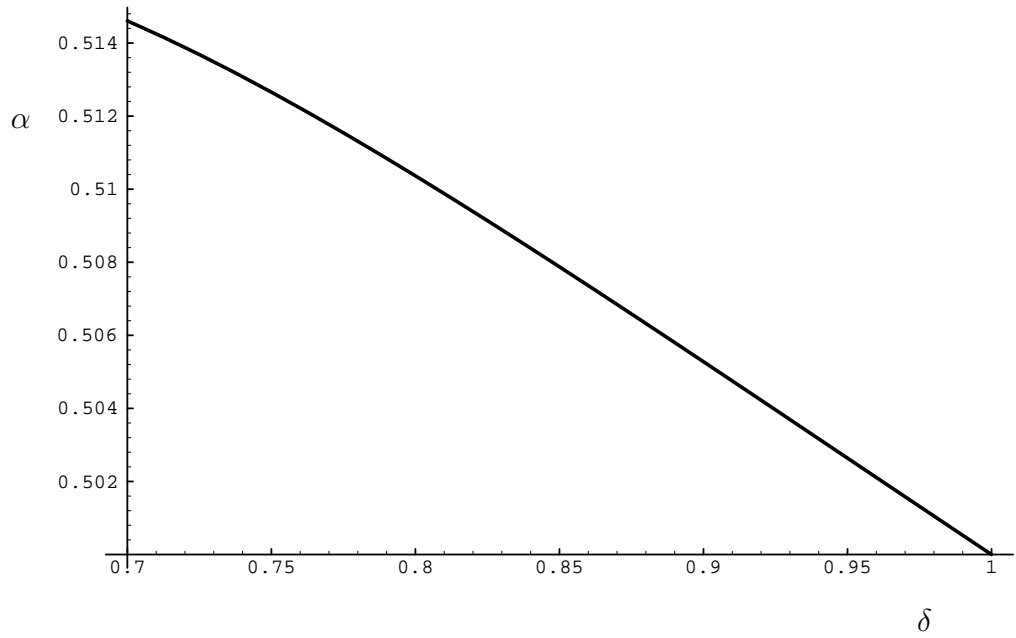}}
\end{center}
\caption{The one loop density decay exponent $\alpha$ for the minority
A species ($\langle a\rangle =O(t^{-\alpha})$) as a function of
$\delta$ (for $0.7\leq\delta\leq 1$).}
\begin{center}
\vspace{-6.35in}
\
\end{center}
\quad $\alpha$
\begin{center}
\vspace{2.15in}
\qquad\qquad\qquad\qquad\qquad\qquad\qquad\qquad\qquad\qquad\qquad
\qquad\quad $\delta$
\end{center}
\begin{center}
\vspace{4.2in}
\
\end{center}
\end{figure}
\begin{figure}
\begin{center}
\leavevmode
\vbox{
\epsfxsize=5in
\epsffile{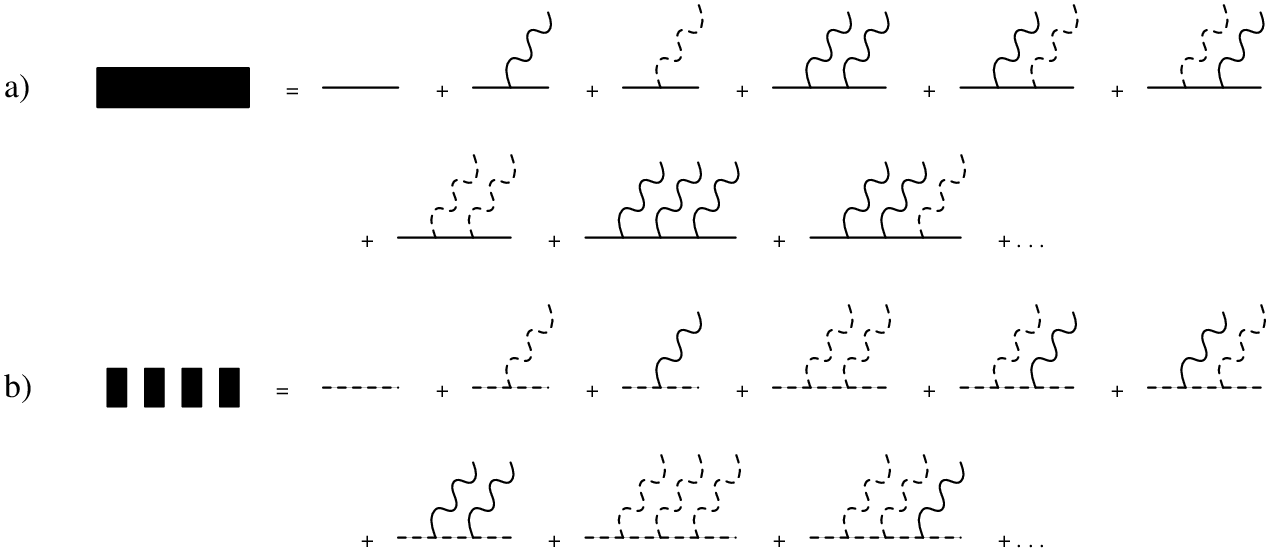}}
\end{center}
\caption{The diagrammatic equations for (a) $\xi$ and (b) $\theta$.}
\end{figure}
\begin{figure}
\begin{center}
\leavevmode
\vbox{
\epsfxsize=5in
\epsffile{figa2.eps}}
\end{center}
\caption{The full diagrammatic equations satisfied by the response
functions.}
\end{figure}
\begin{figure}
\begin{center}
\leavevmode
\vbox{
\epsfysize=2in
\epsffile{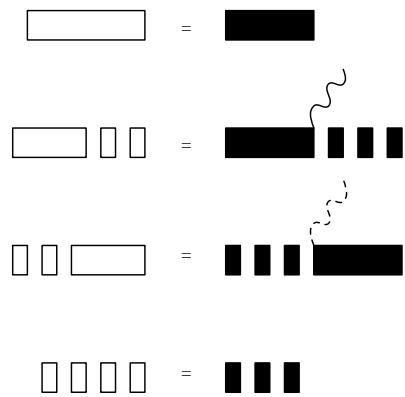}}
\end{center}
\caption{The truncated diagrammatic equations for the response
functions, valid for $\langle a\rangle\gg \langle b\rangle$, or
$\langle b\rangle\gg \langle a\rangle$.}
\end{figure}
\end{document}